\documentclass[10pt, twocolumn, amsmath, amssymb, aps, prx, superscriptaddress]{revtex4-2}
\usepackage{graphicx}
\usepackage{hyperref}
\hypersetup{colorlinks=true,linkcolor=blue}
\usepackage{dcolumn}
\usepackage{physics}
\usepackage{bbold}
\usepackage{units}

\newcommand{\vket}[1]{\ket{#1}\rangle}

\newcommand{\vbraket}[2]{\langle\braket{#1}{#2}\rangle}

\begin{document}

\title{Quantum geometric protocols for fast high-fidelity adiabatic state transfer}

\author{Chris Ventura Meinersen}
\author{Stefano Bosco}
\author{Maximilian Rimbach-Russ}
 
\affiliation{QuTech, Delft University of Technology, Lorentzweg 1, 2628 CJ Delft, The Netherlands
}

\date{\today}

\begin{abstract}
    Efficient control schemes that enable fast, high-fidelity operations are essential for any practical quantum computation. However, current optimization protocols are intractable due to stringent requirements imposed by the microscopic systems encoding the qubit, including dense energy level spectra and cross talk, and generally require a trade-off between speed and fidelity of the operation. Here, we address these challenges by developing a general framework for optimal control based on the quantum metric tensor. This framework allows for fast and high-fidelity adiabatic pulses, even for a dense energy spectrum, based solely on the Hamiltonian of the system instead of the full time evolution propagator and independent of the size of the underlying Hilbert space. Furthermore, the framework suppresses diabatic transitions and state-dependent crosstalk effects without the need for additional control fields. As an example, we study the adiabatic charge transfer in a double quantum dot to find optimal control pulses with improved performance. We show that for the geometric protocol, the transfer fidelites are lower bounded $\mathcal{F}>\unit[99]{\%}$ for ultrafast $\unit[20]{ns}$ pulses, regardless of the size of the anti-crossing. 
\end{abstract}

\maketitle

\section{Introduction}

Coherent control of quantum information is the central part of the advancement of emerging quantum technologies such as quantum processors, quantum sensors, and quantum communication~\cite{glaserTrainingSchrodingerCat2015}. However, the inherently fragile nature of quantum states makes their coherent control a challenging task. Much research is dedicated to finding so-called quantum optimal control protocols, that allow fast and high-fidelity operations by appropriately shaping the control pulses~\cite{poggialiOptimalControlOneQubit2018, liOptimalControlQuantum2023}. Optimized initialization and readout protocols are of particular interest, as they are an integral part of any error correction algorithm~\cite{lidarQuantumErrorCorrection2013}.

\begin{figure*}
    \centering
    \includegraphics[width=\textwidth]{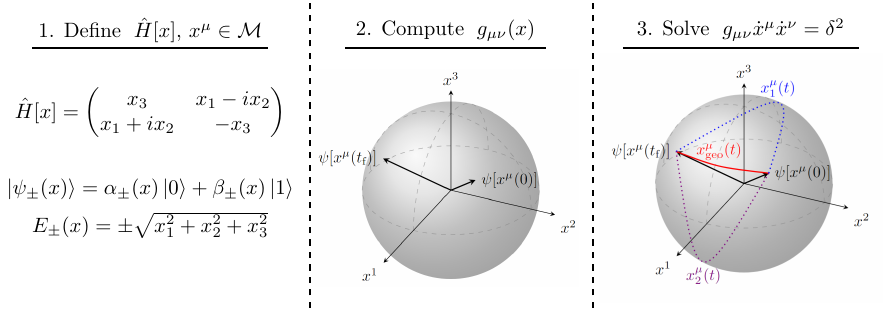}
    \caption{Schematic representation of the geometric fast-QUAD approach with the example of a qubit Hamiltonian. First, one defines the system Hamiltonian and the respective parameters $x^\mu=(x^1,x^2,x^3)$ one wants to optimally operate. The number of parameters defines the dimension of the quantum metric tensor $g_{\mu \nu}$. For a qubit Hamiltonian, the underlying parameter manifold is the Bloch sphere. The geodesic on the sphere (red line) gives the optimal path connecting two states. Examples of non-optimal state transfer protocols ($x^\mu_1,x^\mu_2$) are drawn as blue and purple, dotted lines.}
    \label{fig: geometric flowchart}
\end{figure*}

To achieve a fast and high-fidelity protocol, one has to carefully compose pulse shapes to avoid undesired transitions, which are summarized in shortcut-to-adiabaticity methods~\cite{degrandiAdiabaticPerturbationTheory2010,vuthaSimpleApproachLandauZener2010, guery-odelinShortcutsAdiabaticityConcepts2019,zhuangNoiseresistantLandauZenerSweeps2022,barnesDynamicallyCorrectedGates2022,glasbrennerLandauZenerFormula2023}. Through the addition of new control fields, one can suppress these transitions~\cite{ berryTransitionlessQuantumDriving2009, chenLewisRiesenfeldInvariantsTransitionless2011, selsMinimizingIrreversibleLosses2017, HamiltonianEngineeringAdiabatic, guery-odelinShortcutsAdiabaticityConcepts2019, liuAcceleratedAdiabaticPassage2024}. However, this approach requires additional experimental overhead and precise control of new driving parameters. Approximate methods, based on the minimization of diabatic transitions, circumvent new control fields and only affect the experimentally accessible parameters while providing fast and quasiadabatic (fast-QUAD) protocols~\cite{martinez-garaotFastQuasiadiabaticDynamics2015, xuImprovingCoherentPopulation2019,fehseGeneralizedFastQuasiadiabatic2023}. Unfortunately, these methods cannot be straightforwardly generalized to bigger parameter spaces and beyond transitions between two energy levels. Most protocols are based on classical optimization problems, which are computationally challenging for larger systems~\cite{ khanejaOptimalControlCoupled2005, canevaChoppedRandombasisQuantum2011,goodwinAcceleratedNewtonRaphsonGRAPE2023} and hence make purely numerical methods unattractive. Geometric approaches, including the space curve quantum control~\cite{barnesDynamicallyCorrectedGates2022, zhuangNoiseresistantLandauZenerSweeps2022}, allow for a simple geometric understanding of noisy time dynamics~\cite{walelignDynamicallyCorrectedGates2024}. Notwithstanding, this geometric picture is limited because it relies on the computation of the time evolution operator and is hence constrained to small system sizes~\cite{barnesDynamicallyCorrectedGates2022, zhuangNoiseresistantLandauZenerSweeps2022}. In addition, the derived pulse shapes are natively discontinuous due to the closed-curve and closed-area constraints of the formalism. Similar constraints on the control field to suppress first-, and second-order errors can also be found in~\cite{ribeiroSystematicMagnusBasedApproach2017}.

In this work, we develop a general geometric approach to provide a general framework, based on geodesics provided by the quantum metric tensor~\cite{lambertClassicalQuantumInformation2023, kolodrubetzGeometryNonadiabaticResponse2017, liskaHiddenSymmetriesBianchi2021, chengQuantumGeometricTensor2013, juarezGeneralizedQuantumGeometric2023, houLocalGeometryQuantum2023, kolodrubetzClassifyingMeasuringGeometry2013, maAbelianNonAbelianQuantum2010}, that can be generalized to any multi-level Hamiltonian and allows for fast and high-fidelity adiabatic operations. We refer to this approach as the \textit{geometric fast-QUAD}. Our geometric fast-QUAD does not require imposing any new external control fields like in the counter-diabatic approach, it is resistant to miscalibration errors and allows for fast operations even in dense energy landscapes. It reduces undesired level transitions and the state-dependent crosstalk in the qubit operation and initialization/readout phases. In addition, since it does not require the computation of the full time-ordered evolution operator, we can easily adjust it to allow for operational flexibility of different quantum platforms. We show the advantages of the geometric fast-QUAD through an optimal protocol for initialization and readout of semiconductor spin-qubits.

Semiconductor spin qubits are a platform for quantum computing based on confined semiconductor quantum dots with a promise to be scalable~\cite{scappucciGermaniumQuantumInformation2021, burkardSemiconductorSpinQubits2023}, with long coherence times~\cite{stanoReviewPerformanceMetrics2022}, operability at high temperatures~\cite{undsethHotterEasierUnexpected2023,huangHighfidelitySpinQubit2024}, and their similarity in fabrication to the classical semiconductor industry~\cite{lanzaYieldVariabilityReliability2020, zwerverQubitsMadeAdvanced2022,  cifuentesBoundsElectronSpin2024}. However, their small size can lead to dense energy spectra that may hinder fast and high-fidelity quantum control. For reading out and initializing such qubits, the measured signal is typically enhanced through spin-to-charge conversion techniques~\cite{laiPauliSpinBlockade2011,niegemannParitySingletTripletHighFidelity2022}. Common spin-to-charge techniques, such as Pauli-Spin-Blockade (PSB), rely on an adiabatic transition between a spin and a charge state~\cite{seedhousePauliBlockadeSilicon2021} passing through multiple anticrossings. We illustrate  advantages of the geometric fast-QUAD through optimizing the PSB initialization and readout in a double quantum dot (DQD) with experimentally feasible parameters.

 The paper is structured as follows. In Section~\ref{sec: fq dynamics as geodesics}, the general framework is introduced, starting with the geometric formalism, relating optimal protocols to geodesics, and applying it to a general qubit Hamiltonian. Subsequently, in Section~\ref{sec: dqd model}, upon reviewing the general DQD model in the presence of strong spin-orbit interaction~\cite{hendrickxFourqubitGermaniumQuantum2021}, an effective three-level Hamiltonian describing the readout and initialization subject to PSB is introduced. In addition, we include decoherence sources to provide a detailed analysis of the geometric fast-QUAD.

\section{Quantum geometric formalism: Fast-quasiadiabatic dynamics as geodesics}
\label{sec: fq dynamics as geodesics}

\begin{figure*}
    \centering
    \includegraphics[width=\textwidth]{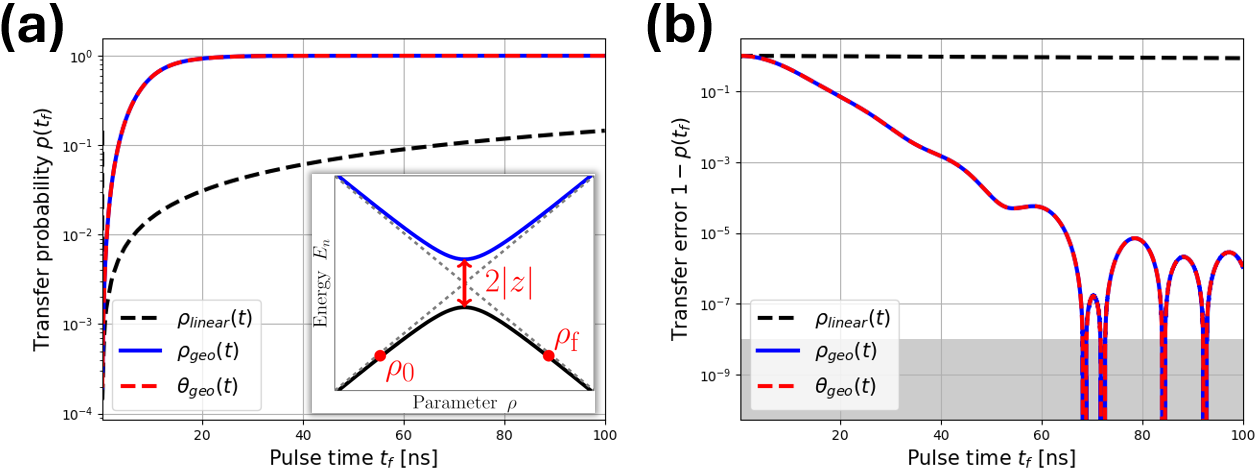}
    \caption{\textbf{(a)} Coherent transfer probability $p(t_\text{f})=|\braket{\psi_0(t_\text{f})}{\psi(t_\text{f})}|^2$ (inset shows energy levels as a function of $\rho$)  and  \textbf{(b)} transfer error $1-p(t_\text{f})$  via quantum geometric protocols for a single qubit system (\ref{eqn: Pauli qubit Hamiltonian}) as a function of pulse time $t_\text{f}$. As the quantum metric tensor (\ref{eqn: Pauli quantum geometric tensor}) is independent of $\phi$ we can choose a protocol exclusively for $\rho(t)$ or for the ratio $\theta(t)=\text{arctan2}(\rho(t),z(t))$. Using the quantum adiabatic protocol (\ref{eqn: adiabatic quantum geo protocol}) we find analytically that $\theta_\text{geo}(t)=(\theta_\text{f}-\theta_\text{0})\,t/t_\text{f}+\theta_\text{0}$, where the adiabaticity is given by the ramp rate $\delta=(\theta_\text{f}-\theta_\text{0})/t_\text{f}$. We also simulate the pulse numerically, as indicated by $\rho_\text{geo}(t)$. Remarkably, the two methods align even up to the minimum transfer error. Throughout these simulations the values of $\phi=0$, $\rho_\text{0}=-10$ GHz, $\rho_\text{f}=10$ GHz, $z=0.1$ GHz have been used. Note that the precision of the numerical simulations determines the singular behavior (gray area) in the transfer error.}
    \label{fig: qugeo single qubit}
\end{figure*}

\subsection{The quantum metric}
Optimal control schemes rely on the control of parameters $x^\mu=(x^1,x^2\dots, x^n)$ of the physical Hamiltonian to provide high-fidelity state transfer. The task of optimizing the fidelity of state transfer can be captured in the geometric structure of the Hilbert space through the quantum metric tensor~\cite{chengQuantumGeometricTensor2013}. The quantum metric tensor  $g_{\mu \nu}$ describes the infinitesimal distance between two pure states via the local infidelity (up to second order in parameter changes)~\cite{tomkaGeodesicPathsQuantum2016,liskaHiddenSymmetriesBianchi2021}
\begin{align}
    \label{eqn: fidelity susceptibility}
    1-|\braket{\psi(x)}{\psi(x+dx)}|^2\approx \frac{1}{2}g_{\mu \nu}(x)dx^\mu dx^\nu.
\end{align} 
Here, $x^\mu \in \mathcal{M}$, where $\mathcal{M}$ is the set of all possible parameter values, define a set of parameters that define an embedding for the set of pure states $P(\mathcal{H})=\mathcal{H}/U(1)$. For Greek indices, we will opt for the Einstein summation convention. The quantum metric tensor constitutes the real and symmetric part of the full quantum geometric tensor $q_{\mu \nu}=g_{\mu \nu}+i\Omega_{\mu \nu}$, whose antisymmetric component $\Omega_{\mu \nu}$ is related to the Berry curvature, which captures topological effects \cite{xiaoBerryPhaseEffects2010, alvarez-jimenezQuantumInformationMetric2017, gianfrateMeasurementQuantumGeometric2020, yuExperimentalMeasurementQuantum2020} allowing for a possibility to straightforwardly connect quantum dynamics and topology. The quantum metric tensor, with respect to a given target state, $\ket{\psi_0}$, can conveniently be written in terms of the Hamiltonian $\hat{H}$, its eigenvalues $E_n$, and eigenvectors $\{\ket{\psi_n}\}$
\begin{align}
    \label{eqn: qgt Hamiltonian}
    g_{\mu \nu}=\Re \sum_{n\neq 0} \frac{\mel{\psi_0}{\partial_\mu \hat{H}}{\psi_n}\mel{\psi_n}{\partial_\nu \hat{H}}{\psi_0}}{(E_n-E_0)^2}=\Re q_{\mu \nu},
\end{align}
where $\partial_\mu = \partial/\partial x^\mu$ is the derivative with respect to the parameters of the Hamiltonian.

\subsection{Minimal energy fluctuations and geodesic equations}
The quantum metric tensor $g_{\mu\nu}$ allows us now to connect fast and quasi-adiabatic (fast-QUAD) dynamics with the geometry of the parameter space. For coherent population transfer, experimentally controlled parameters can be written as $x^\mu(t)$. State transfer is then given by a path connecting the set of initial parameter values $x^\mu_\text{i}\equiv x^\mu(0)$ to some final set $x^\mu_\text{f}\equiv x^\mu(t_\text{f})$. The task of fast and high-fidelity population transfer then relates to the optimization problem of finding an optimal path $x^\mu_\text{geo}(t)$ between these two points described by the following functional
\begin{align}
    \label{eqn: length functional}
    \mathcal{L}[x,\dot{x},t]=\int_{0}^{t_\text{f}} dt \,\sqrt{g_{\mu \nu}(x)\dv{x^\mu}{t}\dv{x^\nu}{t}}.
\end{align}
This functional describes the length of a path $x^\mu(t)$ parametrized by time $t$ and can be minimized for functions $x^\mu_\text{geo}(t)$ (See Fig.~\ref{fig: geometric flowchart}) that solve the Euler-Lagrange equations, which in this context are known as the geodesic equations. For adiabatic protocols, the geodesics have a conserved quantity, namely the total energy, which leads to the geometric adiabatic condition (See Appendix~\ref{app: fundamentals})
\begin{align}
    \label{eqn: adiabatic quantum geo protocol}
    g_{\mu \nu}(x)\dv{x^\mu}{t}\dv{x^\nu}{t}=\delta^2\ll 1.
\end{align}
Here $\delta$ can be interpreted as the adiabaticity parameter. Since the above relationship minimizes the local infidelity~\eqref{eqn: fidelity susceptibility}, the geodesics also minimize the energy fluctuations \cite{bukovGeometricSpeedLimit2019} 
\begin{align}
    \sigma^2_E=\expval{\hat{H}^2}-\expval{\hat{H}}^2\approx g_{\mu \nu}(x)\dv{x^\mu}{t}\dv{x^\nu}{t}
\end{align}
motivating the name of the adiabaticity $\delta=\sqrt{\sigma^2_E}$. If we restrict ourselves to a single parameter $x^\mu=\varepsilon(t)$, we can solve for the adiabaticity parameter as follows
\begin{align}
    \label{eqn: adiabaticity defintion}
    \delta = \frac{1}{t_\text{f}}\int_{\varepsilon(0)}^{\varepsilon(t_\text{f})}d\varepsilon \, \sqrt{g_{\varepsilon \varepsilon}}=\frac{\mathcal{L}[\varepsilon]}{t_\text{f}}\ll 1.
\end{align}
Therefore, adiabatic protocols can be understood as paths that minimize locally the length of the path that they trace out, i.e. short geodesics with respect to the time $t_\text{f}$. The above equation also converges to the quantum speed limit bound for pure states as found in~\cite{bukovGeometricSpeedLimit2019}. Finally, this allows us to draw a connection between adiabatic dynamics and geometry. To find an optimal time evolution of $\varepsilon(t)$, we need to solve
\begin{align}
    \label{eqn: fQ equation}
    g_{\varepsilon \varepsilon}\dot{\varepsilon}^2=\sum_{n\neq 0}\frac{|\mel{\psi_0}{\partial_\varepsilon \hat{H}}{\psi_n}|^2}{(E_n-E_0)^2}\left(\dv{\varepsilon}{t}\right)^2=\delta^2.
\end{align}
Unsurprisingly, this is similar to the known fast-QUAD equation~\cite{martinez-garaotFastQuasiadiabaticDynamics2015, fehseGeneralizedFastQuasiadiabatic2023, xuImprovingCoherentPopulation2019}, differing from the historical fast-QUAD equation only by an additional exponent of 2 of the energy splitting in the denominator~\cite{chenSpeedingQuantumAdiabatic2022}. In contrast, however, it allows for a clear extension to multiple energy levels. In the remainder of the article, we refer to Eq.~\eqref{eqn: fQ equation} as \textit{geometric fast-QUAD} equation. 

\subsection{Two-level system}
The geometric structure of Hilbert space allows us to optimize adiabatic population transfer. For instance, a two-level Hamiltonian in cylindrical coordinates $(\rho, \phi, z)$
\begin{align}
    \label{eqn: Pauli qubit Hamiltonian}
    \hat{H}_\text{Pauli}=\begin{pmatrix}
        z & \rho \, e^{-i\phi} \\
        \rho \, e^{i\phi} & -z 
    \end{pmatrix},
\end{align}
leads to the quantum metric tensor resembling the Bloch sphere (see Appendix~\ref{app: fundamentals})
\begin{align}
    \label{eqn: Pauli quantum geometric tensor}
    [g_{\mu \nu}(\theta,\phi)]= \frac{1}{4}\begin{pmatrix}
        1 & 0 \\
        0 & \sin^2 \theta 
    \end{pmatrix}.
\end{align}
Here $\theta=\text{arctan2}(\rho,z)$ describes the azimuthal angle of the Bloch sphere. Figure \ref{fig: qugeo single qubit} shows the simulated probability $p(t_\text{f})=|\braket{\psi_0(t_\text{f})}{\psi(t_\text{f})}|^2$ under adiabatic evolution for the standard linear protocol $\rho_\text{linear}(t)=(\rho_\text{f}-\rho_\text{0})\,t/t_\text{f}+\rho_\text{0}$ and the geometric protocol as defined in Eq.~\eqref{eqn: adiabatic quantum geo protocol}. Using the quantum adiabatic protocol (\ref{eqn: adiabatic quantum geo protocol}) we find analytically that~\cite{tomkaGeodesicPathsQuantum2016} 
\begin{align}
    \theta_\text{geo}(t)=(\theta_\text{f}-\theta_0)\,t/t_\text{f}+\theta_\text{0},
\end{align}
where $\theta_\text{0},\theta_\text{f}$ are the initial and final values of $\theta(t)$. Remarkably, our analytic expression $\theta_\text{geo}(t)$ and the fully numerically simulated pulse $\rho_\text{geo}(t)$ cannot be distinguished. In both cases, because of the minimization of the energy fluctuations, the transfer errors arising from undesired diabatic transitions are drastically reduced with respect to the linear protocol. 
The geometric fast-QUAD can be easily extended to an arbitrary multi-level system as illustrated in Eq.~\eqref{eqn: qgt Hamiltonian}. In addition, the quantum metric tensor only scales with the number of control parameters and is hence also useful for large systems, making it a reliable tool for analyzing and optimizing large-scale quantum architectures.

\section{Application: Charge transfer in a double quantum dot}
\label{sec: dqd model}

Given the advantages of the quantum metric tensor, we aim to apply the geometric fast-QUAD for the adiabatic initialization and readout processes. We model an effective model for a double quantum dot (DQD) system that may be used directly for the initialization and readout of singlet-triplet qubits \cite{zhangUniversalControlFour2023, jirovecDynamicsHoleSingletTriplet2022,jirovecSinglettripletHoleSpin2021, niegemannParitySingletTripletHighFidelity2022, wangCompositePulsesRobust2012}. After providing a brief overview of the microscopic model in~\ref{sec: full model}, we investigate a truncated two-level model of the full model in Sec.~\ref{sec:truncated}, to extend the previous result of a two-level Landau-Zener problem to one in the presence of $ST_-$ coupling~\cite{jirovecDynamicsHoleSingletTriplet2022,zhangUniversalControlFour2023, saez-mollejoMicrowaveDrivenSinglettriplet2024a}. Following, in Sec.~\ref{sec:untruncated}, we will introduce a low-dimensional effective DQD model, which captures the spin-to-charge transition, while taking into account the spin state. Using this model, we will aim to provide a detailed analysis and comparison of the geometric fast-QUAD with the linear protocol under coherent and non-unitary noise. In the remaining text, we will work in units of $\hbar$.

\subsection{Full model}
\label{sec: full model}
The results in the single qubit case (Fig.~\ref{fig: qugeo single qubit}) can be extended to the full 6x6 DQD~\cite{geyerAnisotropicExchangeInteraction2024, ungererStrongCouplingMicrowave2024}, consisting of two spins in the lowest orbitals of two quantum dots. The Hamiltonian is a sum of the Fermi-Hubbard Hamiltonian and the Zeeman Hamiltonian
\begin{align}
    \hat{H}_{\text{DQD}}=\hat{H}_\text{FH}+\hat{H}_\text{Zeeman},
\end{align}
where the spin-degenerate part is described by the Fermi-Hubbard model
\begin{multline}
\hat{H}_\text{FH} = -\Omega\sum_{ij,\sigma}\Big(\hat{c}_{i,\sigma}^\dagger \hat{c}_{j,\sigma} +\text{h.c.}\Big) +\sum_{\langle ij\rangle}U_{ij}\hat{n}_i\hat{n}_j\\
+\sum_j \left(\frac{\Tilde{U}}{2}\hat{n}_j(\hat{n}_j-1)+V_j\hat{n}_j\right)
\end{multline}
where $\hat{c}^\dagger_{j,\sigma}(\hat{c}_{j,\sigma})$ creates (annihilates) a fermion on sites $j$ with spin $\sigma$. The fermionic number operator is $\hat{n}_j=\sum_\sigma \hat{c}^\dagger_{\sigma,j}\hat{c}_{\sigma,j}$, $\Tilde{U}$ and $U_{ij}$ are the intra- and inter-dot Coulomb repulsion, $\Omega$ is the tunnel coupling originating from the overlap of the wavefunctions in nearby quantum dots, and $V_i$ are the chemical potentials in each dot. The spin degeneracy is lifted through the Zeeman term
\begin{align}
    \hat{H}_\text{Zeeman} &= \frac{1}{2}\mu_B \sum_j \vec{\mathcal{B}}^j \cdot \vec{\sigma}^j,
\end{align}
where $\mu_B$ is the Bohr magneton, $\vec{\sigma}=(\sigma_x,\sigma_y,\sigma_z)^T$ is the Pauli vector consisting of the conventional Pauli matrices, $\mathcal{B}^j_a=\sum_b \mathcal{G}^j_{ab} B_b$ the effective magnetic field and $\mathcal{G}^j_{ab}$ the g-tensor at site $j$. The indices $a, b =x,y,z$ run over the spatial components. Additionally, we define $E_a=E_{a,1}+E_{a,2}$ and $\Delta E_a=E_{a,1}-E_{a,2}$  as the total Zeeman energy and the Zeeman splitting difference, which may arise due to a spatially varying g-factor as usually found in germanium-based platforms \cite{vanriggelen-doelmanCoherentSpinQubit2023, vanriggelenPhaseFlipCode2022, boscoFullyTunableLongitudinal2022, hsiaoExcitonTransportGermanium2024, geyerTwoqubitLogicAnisotropic2022, saez-mollejoMicrowaveDrivenSinglettriplet2024a} or a magnetic field gradient as appears in silicon-based architectures with an additional micromagnet \cite{seedhousePauliBlockadeSilicon2021, takedaQuantumErrorCorrection2022, xueQuantumLogicSpin2022, desmetHighfidelitySinglespinShuttling2024}. The matrix representation in the full basis is given in Appendix~\ref{app: derivation Heff}.

\begin{figure}
    \centering
    \includegraphics[width=0.48\textwidth]{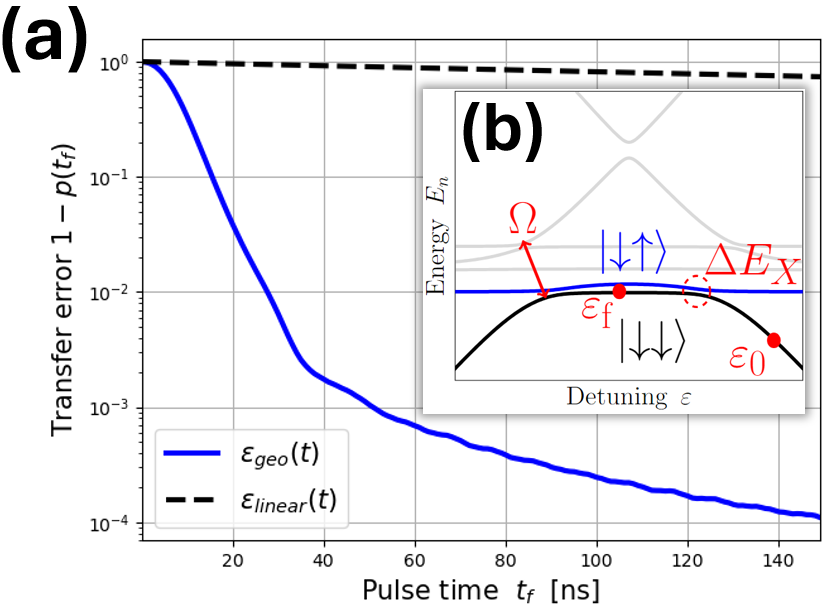}
    \caption{\textbf{(a)} Transfer errors as a function of the pulse time $t_\text{f}$ are plotted for the linear and geometric protocols following Eq.~\eqref{eqn: 6x6 Hamiltonian} without incoherent errors. \textbf{(b)} Illustration of the 6x6 energy levels as a function of detuning, where we only focus on the two lower levels. The $ST_-$ anti-crossing, which induces diabatic transitions in the initialization process, is given by the size of $\Delta E_X= \unit[0.1]{GHz}$, similar to~\cite{saez-mollejoMicrowaveDrivenSinglettriplet2024a} for out-of-plane magnetic fields. After around $\unit[150]{ns}$ pulse time, we can expect an initialization fidelity of $\unit[99.99]{\%}$ for the $\ket{\downarrow\downarrow}$ state. The parameters used in the simulation are: $\Tilde{U}=\unit[100]{GHz}$,  $E_Z=\unit[10]{GHz}$, $\Omega=\unit[10]{GHz}$, $\Delta E_Z=\unit[1]{GHz}$, $\varepsilon_\text{f}=\unit[10]{GHz}$, and $\varepsilon_0=3\Tilde{U}/2$.}
    \label{fig: 6x6 coherent error}
\end{figure}

\subsection{Truncated two-level model}
\label{sec:truncated}

Here, we restrict our optimization protocol to suppress only the transition of the ground state to the closest state. This way we effectively work in a low-energy two-dimensional subspace of the full 6x6 DQD, which in the eigenbasis takes the schematic form
\begin{align}
    \hat{H}_\text{DQD}\approx \sum_{n,m = 0,1}H_{n,m}\ketbra{\psi_n}{\psi_m}.
\end{align}
The state $\ket{T_-}=\ket{\downarrow\downarrow}$ is initialized by shifting the detuning $\varepsilon(t)$, which is the difference of the left and right chemical potentials of each dot $\varepsilon:=V_L-V_R$. Due to the small $ST_-$ anti-crossing, as found in~\cite{saez-mollejoMicrowaveDrivenSinglettriplet2024a} for out-of-plane magnetic fields, we again find that the geometric fast-QUAD is superior to the linear pulse (See Fig.~\ref{fig: 6x6 coherent error}). Even under this simplification, we report a transfer fidelity of $>\unit[99.99]{\%}$ after around $\unit[150]{ns}$ pulse time.

\begin{figure*}
    \centering
    \includegraphics[width=\textwidth]{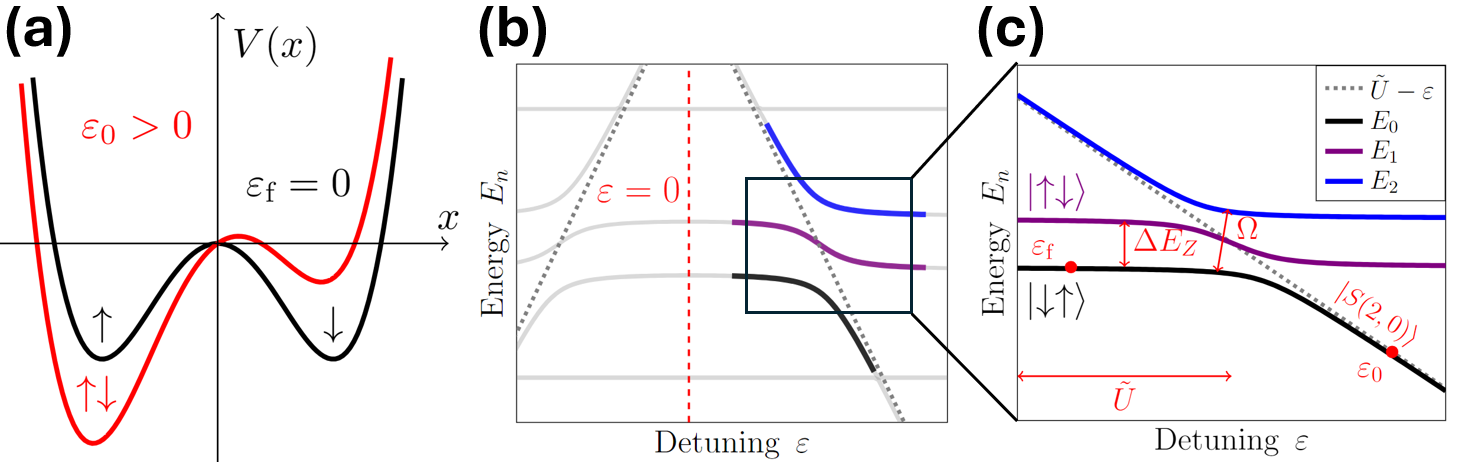}
    \caption{\textbf{(a)} Illustration of the spin-to-charge conversion protocol of spin qubits. The particle on the right side of the DQD potential can only be transferred to an antisymmetric spin-state on the left side due to the Pauli exclusion principle. \textbf{(b)} Plot of the energy spectrum $E_n$ of the full model~\eqref{eqn: 6x6 Hamiltonian} as a function of the detuning $\varepsilon$ (tilt of the potential wells). The colored lines correspond to the energy levels of the reduced three-level model~\eqref{eqn: 3x3 Hamiltonian}. \textbf{(c)} Detailed energy spectrum for the three-level Hamiltonian~\eqref{eqn: 3x3 Hamiltonian}. The dashed line represents the energy of the $\ket{S(2,0)}$ state, which for large detuning $\varepsilon\gg \Tilde{U}, \Omega, \Delta E_Z$ becomes the eigenstate of the DQD Hamiltonian.}
    \label{fig: dqd potential}
\end{figure*}

\begin{figure*}
    \centering
    \includegraphics[width=.85\textwidth]{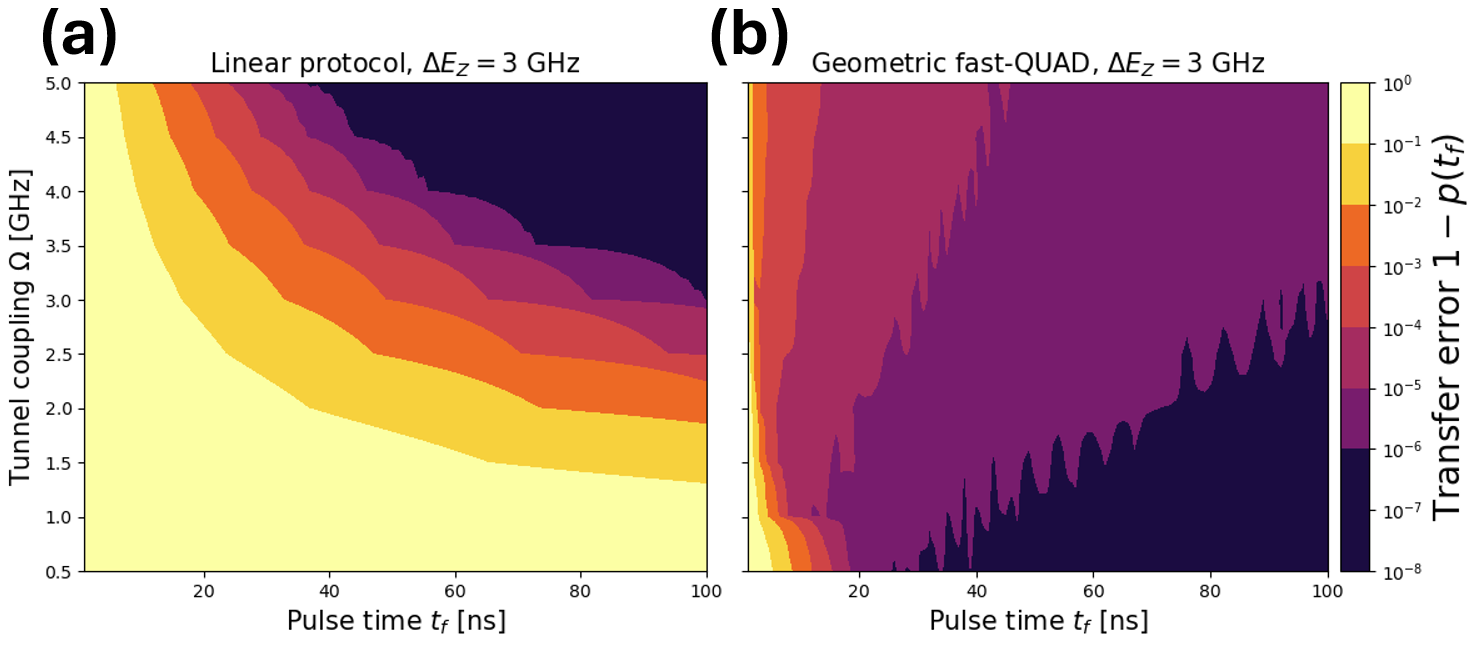}
    \includegraphics[width=.85\textwidth]{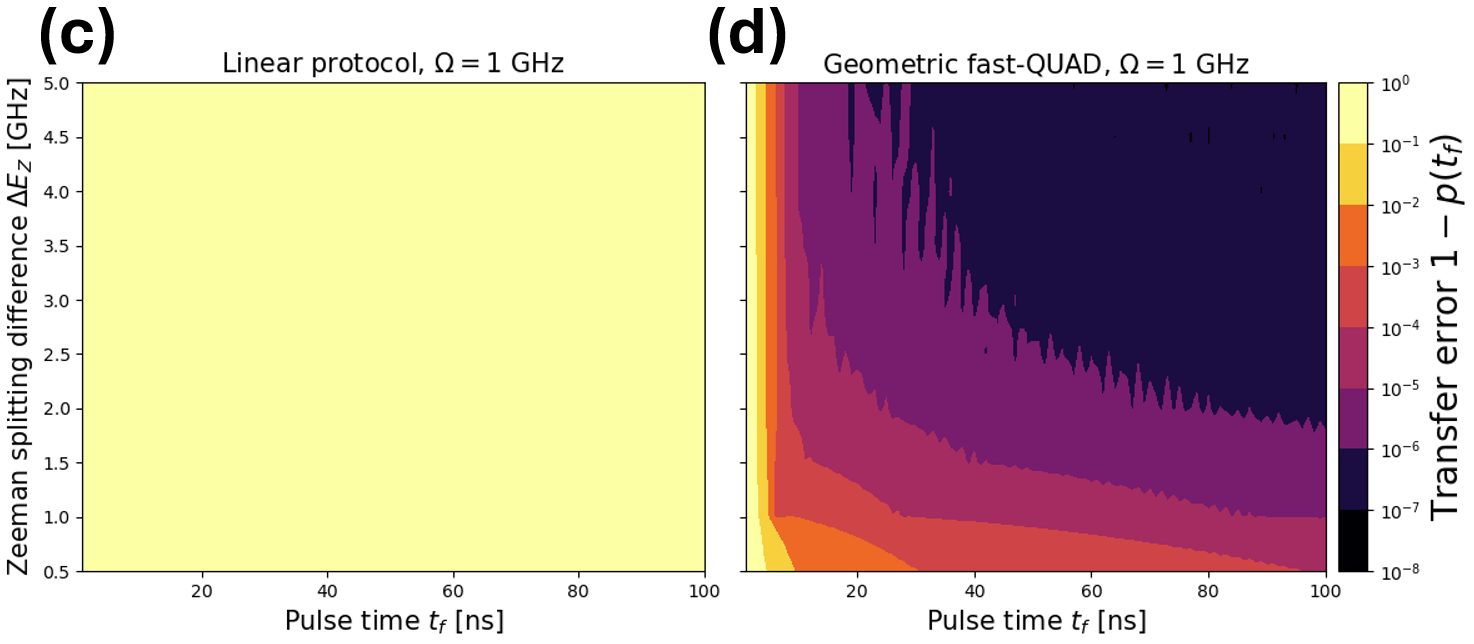}
    \includegraphics[width=.85\textwidth]{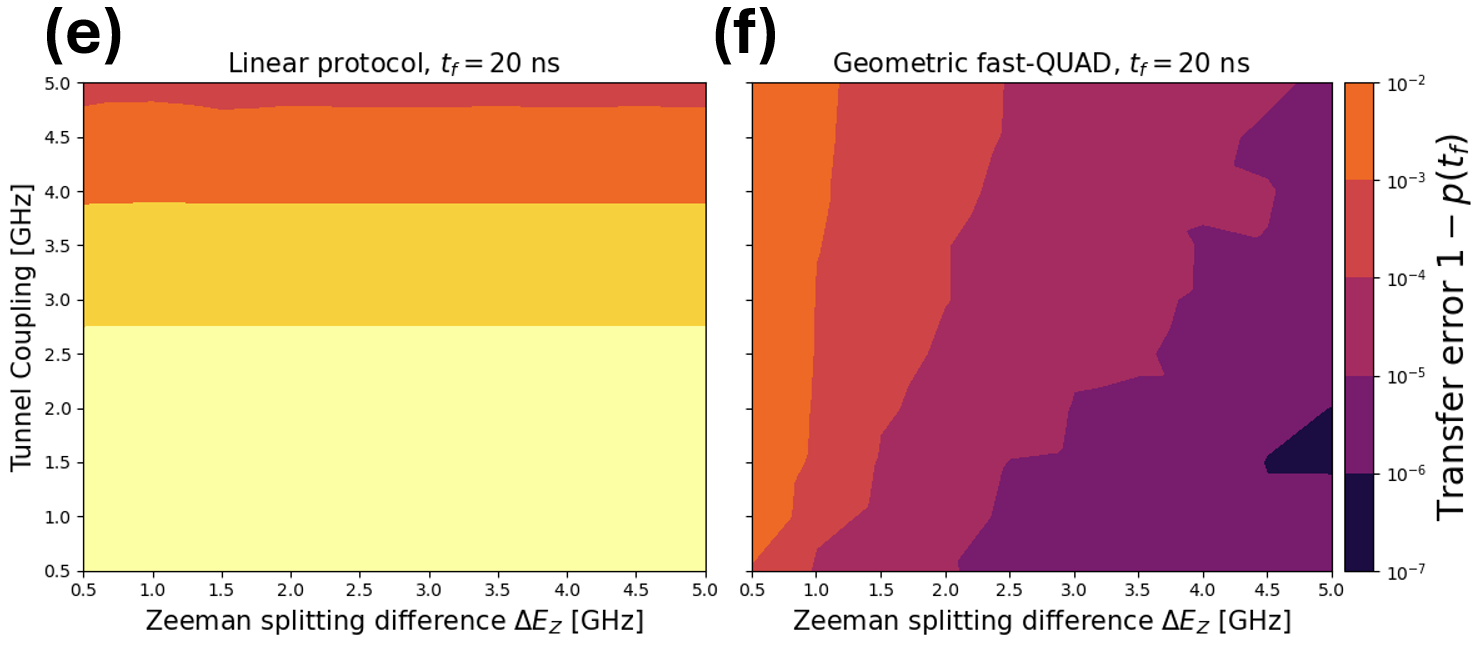}
    \caption{Comparison of initialization pulses using the linear \textbf{(a), (c), (d)} and geometric fast-QUAD \textbf{(b), (d), (f)} protocols  with  $\varepsilon_\text{0}=2\Tilde{U}=200$ GHz and $\varepsilon_\text{f}=0$. \textbf{(a), (b)} Illustrates the effect of both the tunnel coupling and the pulse time on the transfer error. We note that the geometric fast-QUAD achieves a transfer fidelity of $\unit[99.99]{\%}$ after $\unit[20]{ns}$, which defines a lower bound. Transfer errors as low as $10^{-8}$ may be reached even for very small tunnel couplings. \textbf{(c), (d)} is a contour plot sweeping the Zeeman splitting difference and the pulse time $(\Delta E_Z, t_\text{f})$ at fixed tunnel coupling $\Omega=\unit[1]{GHz}$. Here the clear advantage of the geometric fast-QUAD comes through, as the size of the anti-crossing hinders the coherent charge transfer in the linear case. A similar trend can be seen in \textbf{(e), (f)}. Remarkably, in \textbf{(f)} the charge transfer fidelity is lower bounded at $\unit[99]{\%}$, independent of the combination $(\Delta E_Z,\Omega)$ for an ultrafast pulse with pulse time $t_\text{f}=\unit[20]{ns}$. In addition, due to the conservation of energy along adiabatic paths, we observe no difference between the initialization and readout protocols.} 
    \label{fig: coherent error density plot}
\end{figure*}

\subsection{Three-level model}
\label{sec:untruncated}
To obtain a simplified model that captures spin-to-charge conversion (See Fig.~\ref{fig: dqd potential}(a)), we restrict ourselves to a double dot system with a magnetic field pointing purely in the $z$-direction, neglecting the fully polarized states $\ket{T_\pm}$, and focus on the singlet-triplet basis with two charge degrees of freedom (Fig.~\ref{fig: dqd potential}(b)). The Hilbert space is spanned by the following basis states 
\begin{align}
    \ket{S(2,0)} &=\hat{c}^\dagger_{L,\uparrow}\hat{c}^\dagger_{L,\downarrow} \ket{\text{vac}} \\
    \ket{S(1,1)} &=\frac{1}{\sqrt{2}}\left(\hat{c}^\dagger_{L,\uparrow}\hat{c}^\dagger_{R,\downarrow}- \hat{c}^\dagger_{L,\downarrow}\hat{c}^\dagger_{R,\uparrow}\right)\ket{\text{vac}} \\
    \ket{T_0(1,1)} &=\frac{1}{\sqrt{2}}\left(\hat{c}^\dagger_{L,\uparrow}\hat{c}^\dagger_{R,\downarrow}+ \hat{c}^\dagger_{L,\downarrow}\hat{c}^\dagger_{R,\uparrow} \right)\ket{\text{vac}}, 
\end{align}
where $(n_L, n_R)$ describes the number of charges in the left and right dots, respectively, and $\ket{\text{vac}}$ represents the vacuum state. In this subspace, we find that the matrix representation of the DQD Hamiltonian, in the above basis, is
\begin{align}
    \label{eqn: 3x3 Hamiltonian}
    \hat{H}(t)=\begin{pmatrix}
        \Tilde{U}-\varepsilon(t) & \Omega & 0\\
        \Omega & 0 & \Delta E_Z\\
        0 & \Delta E_Z & 0\\
    \end{pmatrix}.
\end{align}
The energy spectrum of the Hamiltonian in Eq.~\eqref{eqn: 3x3 Hamiltonian} is seen in Fig.~\ref{fig: dqd potential}(c), which displays an anti-crossing at $\varepsilon=\Tilde{U}$. The size of the anti-crossing is now determined by the combination of the tunnel coupling $\Omega$ and the Zeeman splitting difference $\Delta E_Z$. In contrast to the Landau-Zener-Majorana-Stueckelberg anticrossing, the energy spectrum is not symmetric in the detuning $\varepsilon$. Also, the existence of a third energy level, makes it possible for diabatic transitions from the ground state to two upper energy eigenstates.

\subsection{Incoherent dynamics}
During coherent spin-to-charge conversion, a dominant error source are diabatic transitions in the vicinity of the anti-crossings, where the energy level difference is minimal. Our geometric protocol Eq.~\eqref{eqn: adiabatic quantum geo protocol} is designed to minimize such errors. However, non-unitary dynamics arise due to couplings with ambient degrees of freedom, which alter the time evolution and hence may affect the optimal pulse shape. In the following, we will describe two ubiquitous noise types that may affect the protocol, low- and high-frequency charge noise.

One of the most known types of noise in semiconducting and superconducting devices is the appearance of low-frequency noise \cite{paladinoNoiseImplicationsSolidstate2014, yonedaQuantumdotSpinQubit2018, koganElectronicNoiseFluctuations1996, zouSpatiallyCorrelatedClassical2023, greenArbitraryQuantumControl2013a, duttaLowfrequencyFluctuationsSolids1981, burinManyElectronTheory2006, hansenAccessingFullCapabilities2023}, whose noise spectral density follows $S(f)\propto 1/f$. Under sufficient conditions, the noise spectral density is the Fourier transform of the auto-correlation of the noise. For the Hamiltonian in Eq.~\eqref{eqn: 3x3 Hamiltonian}, the noise spectral density arises from the correlation function of the fluctuations of the detuning parameter $\delta \varepsilon(t)$, which we include as a perturbation 
\begin{align}
    \label{eqn: 1f noise detuning}
    \delta \hat{H}(t)=-\delta \varepsilon(t) \ketbra{S(2,0)}=-\delta\varepsilon(t) \,\hat{\Pi}_\text{S(2,0)},
\end{align}
where $\delta \varepsilon$ is drawn from a Gaussian distribution $\delta \varepsilon\sim \mathcal{N}(0,\sigma^2)$, which is centered at zero with a variance given by $\sigma^2$. In the simulations, we will model this behavior by fluctuating boundary conditions of the pulse shape $\varepsilon_{0,\text{f}}\to \varepsilon_{0,\text{f}}+\delta \varepsilon$~\cite{krzywdaDecoherenceElectronSpin2024} and illustrate the differences between a noisy and a noiseless time evolution.

To study the high-frequency noise, we will adopt the Lindblad master equation, which describes non-unitary evolution of a quantum system subject to Markovian noise. It takes the form
\begin{align}
    \dv{\hat{\rho}}{t}=-i\, [\hat{H},\hat{\rho}]+\sum_j \Big( \hat{L}_j\hat{\rho}\hat{L}_j^\dagger -\frac{1}{2}\{ \hat{L}_j^\dagger \hat{L}_j, \hat{\rho} \}  \Big),
\end{align}
where $\hat{L}_j$ are the conventional Linblad operators. Explicitly, we describe dephasing from high-frequency noise via the dephasing jump operator~\cite{seedhousePauliBlockadeSilicon2021}
\begin{align}
    \hat{L}_\text{dephasing} &=\sqrt{\frac{1}{2T_2}}\begin{pmatrix}
        1 & 0 & 0 \\
        0 & -1 & 0 \\
        0 & 0 & -1 \\
    \end{pmatrix}.
\end{align}
Here the Lindblad operator acts on the charge states (1,1) and (2,0). The strength of the dephasing is captured by the decoherence time $T_2$. Since the dephasing operator has real entries hermitian, we can also shift the Lindbladian to obtain the equivalent dephasing operator
\begin{align}
    \hat{L}_\text{dephasing}'= \hat{L}_\text{dephasing} + \sqrt{\frac{1}{2T_2}}\,\hat{\mathbb{1}}= \sqrt{\frac{2}{T_2}}\,\hat{\Pi}_\text{S(2,0)},
    \label{eqn: jump operators}
\end{align}
which will generate the same time dynamics. Since for spin qubits relaxation is several orders of magnitude longer than dephasing for spin~\cite{stanoReviewPerformanceMetrics2022} and charge qubits~\cite{macquarrieProgressCapacitivelyMediated2020}, we neglect it in our analysis. Note that relaxation usually benefits adiabatic charge transfer of the ground-state as it counteracts the diabatic transitions to energetically excited states, thus effectively improving the population transfer fidelity. 

Using the Uhlmann fidelity $\mathcal{F}$ defined as
\begin{align}
    \mathcal{F}(\rho, \sigma)=\Big( \tr \sqrt{\sqrt{\rho}\sigma \sqrt{\rho}}\Big)^2,
\end{align}
we can determine the overlap of the lowest energy eigenstate $\ket{\psi_0(t)}$ and the time-evolved one under the non-unitary evolution given by~\eqref{eqn: jump operators} for the linear and geometric state transfer protocols.

\begin{figure}
    \centering
    \includegraphics[width=0.48\textwidth]{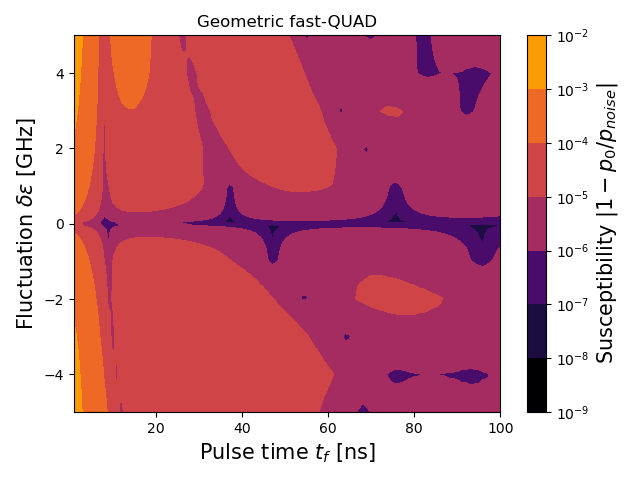}
    \caption{Susceptibility of the geometric fast-QUAD to low-frequency fluctuations in detuning. We simulate the effects of quasistatic noise via a fluctuation in the detuning (\ref{eqn: 1f noise detuning}). This fluctuation will affect the initial and final boundary conditions when simulating the pulse shape as follows: $\varepsilon_{0,\text{f}}\to \varepsilon_{0,\text{f}}+\delta \varepsilon$. We fix the values of the tunnel coupling to $\Omega=\unit[3]{GHz}$, $\Delta E_Z=\unit[0.5]{GHz}$, and the remaining are identical to the ones in Figure \ref{fig: coherent error density plot}. The errors mostly affect infidelities below $1-\mathcal{F}<10^{-4}$ after a pulse time of $t_\text{f}=\unit[20]{ns}$. The asymmetry of the deviation is due to the fact that the energy spectrum is asymmetric around the anti-crossing at $\varepsilon=\Tilde{U}$.}
    \label{fig: 1/f deviation}
\end{figure}

\subsection{Results}
\label{sec: results}

For concreteness, we will focus in our analysis only on the initialization process, as the readout process is directly provided by the reverse pulse shape. The initial state is the singlet state in the (2,0) charge state and is adiabatically pulsed to the desired final state. Under coherent evolution, the only error source is due to undesired diabatic transitions, inducing interference effects that reduce the transfer fidelity. We scan multiple pairs of parameters and simulate the transfer error $1-p(t_\text{f})$ using the protocol in Eq.~\eqref{eqn: fQ equation} and the linear protocol. For the geometric protocol, we generate an appropriate adiabaticity using the boundary conditions of the detuning $(\varepsilon_0,\varepsilon_\text{f})$ and the pulse time $t_\text{f}$ with Eq.~\eqref{eqn: adiabaticity defintion} and then feed the numerically solved pulse $\varepsilon_\text{num}(t)$ into our Hamiltonian or Lindbladian based time evolution, depending on whether we want to study unitary or non-unitary dynamics. 

\paragraph{Unitary dynamics} Figure~\ref{fig: coherent error density plot} shows the results of the geometric pulse for the initialization sequence of the $\ket{\downarrow\uparrow}$ state of the Hamiltonian (\ref{eqn: 3x3 Hamiltonian}). The transfer error is reduced as a function of the pulse time as we move more adiabatic at larger pulse times. We observe the advantage of using the geometric fast-QUAD over the linear protocol as a reliable protocol for circumventing coherent errors, even for very small anti-crossings and extremely fast pulse times. Strikingly, we observe as a common trend, that the transfer fidelity for the geometric fast-QUAD for $t_\text{f}>\unit[20]{ns}$ yields a transfer fidelity $\mathcal{F}>\unit[99]{\%}$ for all investigated settings of tunnel couplings $\Omega$ and Zeeman splitting differences $\Delta E_Z$. Note that these results, for the same parameter settings besides the detuning boundary conditions, also hold for the adiabatic readout protocol, as the energy is conserved along these paths.

\begin{figure}
    \centering
    \includegraphics[width=0.45\textwidth]{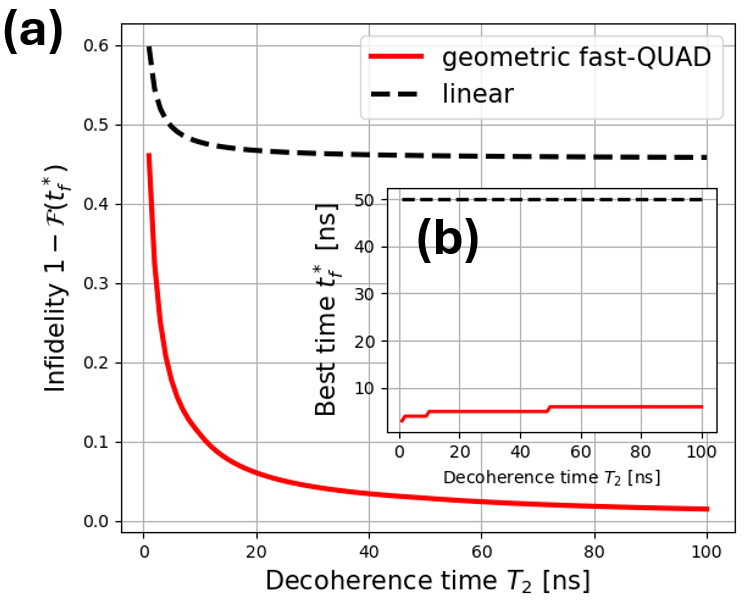}
    \caption{\textbf{(a)} Plot of the minimum infidelity of the geometric fast-QUAD (red line) and linear protocol (black, dashed line) for a given decoherence time $T_2$ for pulses up to $\unit[50]{ns}$. The geometric fast-QUAD, even under dephasing noise and dense energy spectra ($\Omega=\unit[1]{GHz}$, $\Delta E_Z=\unit[1]{GHz}$), does always provide higher fidelity protocols at ultrafast operation times ($t_\text{f}^*<\unit[10]{ns}$). The boundary conditions chosen are: $\varepsilon_\text{0}=2\Tilde{U}=200$ GHz, and $\varepsilon_\text{f}=0$. \textbf{(b)} Best pulse operation times given a decoherence time $T_2$. Note that the linear protocol, due to the fact that it induces diabatic transitions, will always pick the longest operation time to achieve the highest fidelity.}
    \label{fig: infidelity nonunitary}
\end{figure}

\begin{figure}
    \centering
    \includegraphics[width=0.45\textwidth]{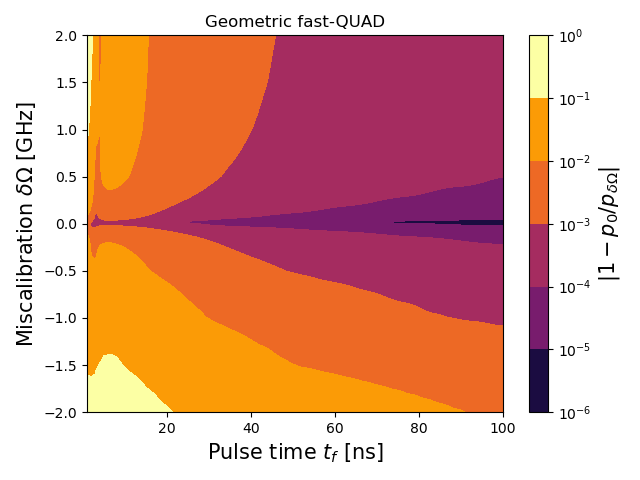}
    \caption{Transfer fidelity deviation, due to a miscalibrated pulse $\Omega_\text{pulse}=\Omega_0 +\delta \Omega$, where the system Hamiltonian is set at $\Omega_0=\unit[3]{GHz}$ and $\Delta E_Z=\unit[0.5]{GHz}$. Remarkably, assuming a bigger tunnel coupling for the pulse will deviate less from the calibrated result.}
    \label{fig: miscalibration plot}
\end{figure}

\paragraph{Non-unitary dynamics} 
In addition to providing higher fidelities for very short pulse times and dense energy spectra, our protocol is also highly resilient against quasistatic noise as seen in Fig.~\ref{fig: 1/f deviation}, where we plotted the susceptibility of the transfer fidelity with respect to detuning fluctuations. Notably, after $\unit[20]{ns}$ the effects of the quasistatic noise affect the fidelities below $10^{-4}$, even for strong fluctuation of $\delta \varepsilon=\unit[5]{GHz}$. In Fig.~\ref{fig: miscalibration plot} we also show that the geometric protocol is robust against pulse miscalibration regarding $\Omega\rightarrow \Omega + \delta \Omega$, for small $\delta \Omega$ and for pulse times bigger than $\unit[20]{ns}$. Remarkably, assuming larger tunnel couplings for the pulse leads to smaller deviations from the calibrated result. This may be caused due to favorable interference effects in the initial ramp towards the anti-crossing. Namely, assuming a smaller anti-crossing will usually result in an initial fast ramping ending in a slow-down at the anti-crossing, leading to interference effects that do not destructively interfere beyond the anti-crossing.

Lastly, we compare the optimal control sequences of the linear and geometric protocol given some fixed decoherence time $T_2$ in Fig.~\ref{fig: infidelity nonunitary}. Here, we compute the linear and geometric fast-QUAD pulse shapes and simulate the Uhlmann fidelity of the lowest energy eigenstate $\ket{\psi_0(t_\text{f})}$ and the resulting mixed density matrix $\rho(t_\text{f})$ under Lindbladian time evolution. For a fixed interval of allowed pulse operation times $t_\text{f}\leq \unit[50]{ns}$, we find the maximum Uhlmann fidelity at fixed decoherence time and compare the linear and geometric protocol. The maximum Uhlmann fidelity $\mathcal{F}(t_\text{f}^*)$ is given by the optimal pulse operation time $t_\text{f}^*$ for any given decoherence time $T_2$. The linear protocol will always perform best at the longest allowed pulse times to suppress the diabatic transitions which dominate in this regime. On the other hand, the geometric fast-QUAD will be optimal at ultrafast operation times ($t_\text{f}^*<\unit[10]{ns}$) to  simultaneousely reduce the coherent and incoherent errors through dephasing. Therefore, the geometric fast-QUAD will always outperform the linear protocol under fixed decoherence rates.

\section{Conclusion}

In this work, we established a relationship between the quantum geometric structure of the Hilbert space and  quasiadiabatic time dynamics. Our main focus lied in providing a general framework to deal with coherent errors arising from undesired diabatic transitions between multiple energy levels while operating at fast pulse times. Special emphasis was put on applying these methods to enable fast and high-fidelity adiabatic initialization and readout for a DQD system, which is integral for the minimization of state-dependent crosstalk. Nevertheless, we stress that the framework is applicable to all quantum systems with non-degenerate eigenvalues and only requires optimization on the level of the Hamiltonian and not the time-ordered evolution operator. For coherent evolution, we found that independent of the parameter configuration, the geometric fast-QUAD provides an upper bound on the transfer error at $10^{-2}$ for pulse times of $\unit[20]{ns}$.  Miscalibration of the pulse and quasistatic noise did not yield significant deviations. Errors arising due to dephasing effects were studied, and it was found that the geometric fast-QUAD was always superior to the linear protocol with respect to the Uhlmann fidelity, while allowing ultrafast operation times.

Nevertheless, further efforts in the understanding of the quantum geometric approach have to be made, especially with a focus on including the effects of noise directly into the formalism, understanding the geometric difference between pure and mixed states, and what the impact of a non-abelian connection and degenerate eigenstates would be for transfer protocols. So far, the case of mixed states has been tackled only for full and finite rank density matrices \cite{houLocalGeometryQuantum2023, hornedalGeometricalDescriptionNonHermitian2024}, making it challenging to capture non-unitary time evolution and systems with infinite-dimensional Hilbert spaces. Nevertheless, the quantum geometry of parameter space will provide new opportunities for ultrafast adiabatic operations allowing for significant improvements in the coherent processing of quantum information, accelerating the advancement of emerging quantum technologies.

\section*{Acknowledgments}

We thank the members of the Veldhorst, Scappucci, and Vandersypen groups for helpful discussions on practical applications. Additionally, we are grateful for discussions with Amanda Seedhouse and Edmondo Valvo about the thoeretical model. This research was partly supported by the EU through the H2024 QLSI2 project and partly sponsored by the Army Research Office under Award Number: W911NF-23-1-0110. The views and conclusions contained in this document are those of the authors and should not be interpreted as representing the official policies, either expressed or implied, of the Army Research Office or the U.S. Government. The U.S. Government is authorized to reproduce and distribute reprints for Government purposes notwithstanding any copyright notation herein.
\appendix

\section{Fundamentals of quantum Riemannian geometry}
\label{app: fundamentals}

\paragraph{Bloch sphere from quantum geometry}

To define the metric $g$ we need to find a basis $t_\mu(x)$ that spans the tangent space $T_{\hat{\rho}} P(\mathcal{H})$. A natural choice is given by the set of traceless and Hermitian matrices

\begin{align}
    t_\mu(x)=\partial_\mu \hat{\rho}(x)=\ketbra{\partial_\mu \psi}{\psi}+\ketbra{\psi}{\partial_\mu \psi},
\end{align}

where we assume that the density matrices are pure $\hat{\rho}(x)=\ketbra{\psi(x)}$ and that the derivative is with respect to the parameters $x^\mu$.  We can define the quantum geometric tensor as the Killing form on the tangent space $T_{\hat{\rho}} P(\mathcal{H})$

\begin{align}
    \label{eqn: metric defintion}
    g_{\mu \nu}&=\frac{1}{2}\tr(t_\mu t_\nu)\\
    &=\Re\Big[\braket{\partial_\mu \psi}{\partial_\nu \psi}\Big]+\braket{\partial_\mu \psi}{\psi}\braket{\psi}{\partial_\nu \psi}.
\end{align}

We note that the quantum geometric tensor (QGT) has certain symmetries, which, in part, will constrain our dynamics. First, the QGT is invariant under shifts in the ground state energy $\hat{H}\to \hat{H}+\omega(x) \hat{\mathbb{1}}$, which is the known invariance that only energy differences are measurable and is the expected invariance under $U(1)$. Secondly, the QGT does not change if we rescale the Hamiltonian globally with a function $\Omega(x)$, i.e. $\hat{H}\to \Omega(x)\hat{H}$, which we will refer to as conformal invariance. We also need to rescale the time variable to not affect the time dynamics. The conformal invariance will constrain our dynamics to a $(\dim \mathcal{M}-1)$-dimensional subspace. To see this, we will work through the example in the main text: A general $2\times 2$ Hamiltonian can be written in the Pauli basis, which in polar coordinates $(\rho, \phi, z)$ takes the form

\begin{align}
    \hat{H}_\text{Pauli}=\begin{pmatrix}
        z & \rho \, e^{-i\phi} \\
        \rho \, e^{i\phi} & -z 
    \end{pmatrix},
\end{align}

where we note that, for pure states, $\dim P(\mathcal{H}_\text{Pauli})=2<\dim \mathcal{M}=3$, as pure states can be fully described by the angles $(\theta, \phi)$ on the Bloch sphere. This condition restricts the notion of $\mathcal{M}$ being an embedding of the projective Hilbert space $P(\mathcal{H}_\text{Pauli})$ as the map is no longer injective. Due to the conformal invariance, however, we may restrict ourselves to subspaces that span the projective Hilbert space and hence form a well-defined embedding. For instance, if we identify $x^\mu =\{\rho, \phi, z\}$ then we may find a function $\Omega(x)$ such that we can reduce the number of parameters. If we want to work in the subspace of $x^\mu =\{\rho,  z\}$ we find that the quantum metric tensor is singular, i.e. $\det g=0$, which alludes to the fact that the embedding is ill-defined. This feature can be seen by the fact that there is no non-trivial function $\Omega(x)$ that removes the $\phi$-dependence. On the other hand, the subsets $x^\mu =\{\rho,  \phi\}$ and $x^\mu =\{\phi,  z\}$  can be well-defined. For instance, if $\Omega(x)=z$ and we redefine $\rho/z\to \rho$, the Pauli Hamiltonian takes the form

\begin{align}
    \hat{H}_\text{Pauli}=\begin{pmatrix}
        1 & \rho \, e^{-i\phi} \\
        \rho \, e^{i\phi} & -1 
    \end{pmatrix},
\end{align}

which leads to a non-singular quantum metric tensor

\begin{align}
    [g_{\mu \nu}(\rho, \phi)]= \frac{1 }{4(1+\rho^2)}\begin{pmatrix}
        \frac{1}{(1+\rho^2)} & 0 \\
        0 & \rho^2 
    \end{pmatrix},
\end{align}

which captures the fact that the embedding is well-defined. This can also be seen by the fact that now $\dim P(\mathcal{H}_\text{Pauli})=\dim \mathcal{M}$. This metric is the metric on the Bloch sphere, as can be seen if we use $\rho=\tan \theta$ and use the transformation rule for the quantum metric tensor to arrive at 

\begin{align}
    [g_{\mu \nu}(\theta,\phi)]= \frac{1}{4}\begin{pmatrix}
        1 & 0 \\
        0 & \sin^2 \theta 
    \end{pmatrix}.
\end{align}

\paragraph{Beltrami identity, Killing charges and energy fluctuations}

To derive that adiabatic geometric condition (\ref{eqn: adiabatic quantum geo protocol}) we start by simplifying the length functional in the main text via the Cauchy-Schwarz relation to the following functional \cite{tomkaGeodesicPathsQuantum2016}

\begin{align}
    \mathcal{L}'[x,\dot{x},t]=\int_{0}^{t_\text{f}} dt \, \Big[g_{\mu \nu}(x)\dot{x}^\mu\dot{x}^\nu\Big],
\end{align}

where the integrand can be understood as a Lagrangian $L[x,\dot{x},t]$ and the functional as the action. If the Lagrangian does not explicitly depend on time, i.e. $\partial L/\partial t=0$, then Beltrami's identity holds

\begin{align}
    \dot{x}^\alpha\pdv{L}{\dot{x}^\alpha}-L=\text{const.}
\end{align}

The left-hand side is the expression of the Hamiltonian and hence, in this case, Beltrami's identity is a consequence of conservation of energy. Computing the partial derivative of the Lagrangian above using $\partial \dot{x}^\mu/\partial \dot{x}^\alpha=\delta^\mu_\alpha$ we find the adiabatic-geometric condition 

\begin{align}
    g_{\mu \nu}(x)\dot{x}^\mu\dot{x}^\nu=\text{const.} 
\end{align}

Another way to see this identity is that conservation laws arise due to symmetries. As we are considering unitary systems, we have time-reversal symmetry and hence energy conservation. The connection between symmetry and conserved charges in the geometrical context is illustrated by the Killing vectors $\xi^\mu$. Each Killing vector has an associated conserved charge \cite{liskaHiddenSymmetriesBianchi2021}
\begin{align}
    \partial_t Q_\xi = \partial_t \Big( g_{\mu \nu}(x) \xi^\mu \dot{x}^\nu \Big)=0.
\end{align}
If the Killing vector is proportional to the tangent vector $\xi^\mu \propto \dot{x}^\mu$ we also find the adiabatic-geometric relation. This shows the explicit relation between energy conservation and geometry. In order to find the geodesics of a manifold one also only needs $\dim \mathcal{M}-1$ Killing vector fields \cite{liskaHiddenSymmetriesBianchi2021}, which aligns with the parameter subspace after the constraint due to the conformal invariance of the QGT.

\begin{figure}
    \centering
    \includegraphics[width=0.45\textwidth]{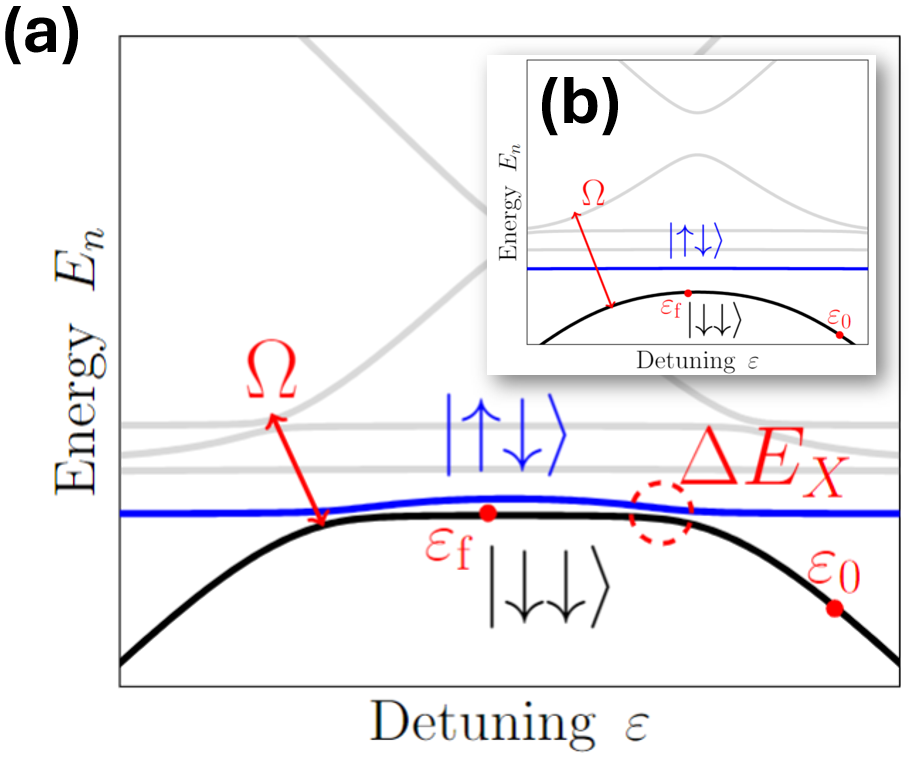}
    \caption{Energy spectrum 6x6 comparison for low \textbf{(a)} and high \textbf{(b)} tunnel couplings. The size of the anticrossing induced by $\Delta E_X$ is removed by the increment of the tunnel coupling. This allows for very fast pulses as one decreases the probability of Landau-Zener transitions. Note that the Zeeman energy is comparable to the tunnel coupling, such that for the large tunnel coupling we still define the $\ket{\downarrow\downarrow}$ as the ground state at zero detuning.}
    \label{fig: energy spectrum 6x6 comparision}
\end{figure}

\section{DQD Eigenspectrum and effective 2x2 Hamiltonian}
\label{app: derivation Heff}

We want to describe the effective dynamics of the DQD Hamiltnoian in Eq.~\eqref{eqn: 3x3 Hamiltonian}. The total Hamiltonian describing an array of quantum dots is given by

\begin{align}
    \hat{H}_{\text{DQD}}=\hat{H}_\text{FH}+\hat{H}_\text{Zeeman},
\end{align}

where these two Hamiltonians can be written as~\cite{ungererStrongCouplingMicrowave2024, geyerAnisotropicExchangeInteraction2024}
\begin{widetext}
    \begin{align}
\label{eqn: 6x6 Hamiltonian}
    \hat{H}_{\text{DQD}} = \begin{pmatrix}
\Tilde{U} + \varepsilon & 0 & 0 & -\Omega & \Omega & 0 \\
0 & \Tilde{U} - \varepsilon & 0 & -\Omega & \Omega & 0 \\
0 & 0 & E_Z & \Delta E_X & -\Delta E_X & 0 \\
-\Omega & -\Omega & \Delta E_X & \Delta E_Z & 0 & \Delta E_X \\
\Omega & \Omega & -\Delta E_X & 0 & -\Delta E_Z & -\Delta E_X \\
0 & 0 & 0 & \Delta E_X & -\Delta E_X & -E_Z
\end{pmatrix}
\end{align}
\end{widetext}
if we constrain ourselves to the charge states (1,1), (2,0), and (0,2) including the spin states $\ket{\uparrow\uparrow},\ket{\uparrow\downarrow},\ket{\downarrow\uparrow},\ket{\downarrow\downarrow}\}$. In addition, we define $E_j=E_{j,1}+E_{j,2}$ and $\Delta E_j=E_{j,1}-E_{j,2}$ with $j=X,Y,Z$ for each component of the Pauli vector, and we set the $y$-component to zero for simplicity. The energy spectrum of the above Hamiltonian is plotted in Figure \ref{fig: energy spectrum 6x6 comparision}.

\begin{figure}
    \centering
    \includegraphics[width=0.5\textwidth]{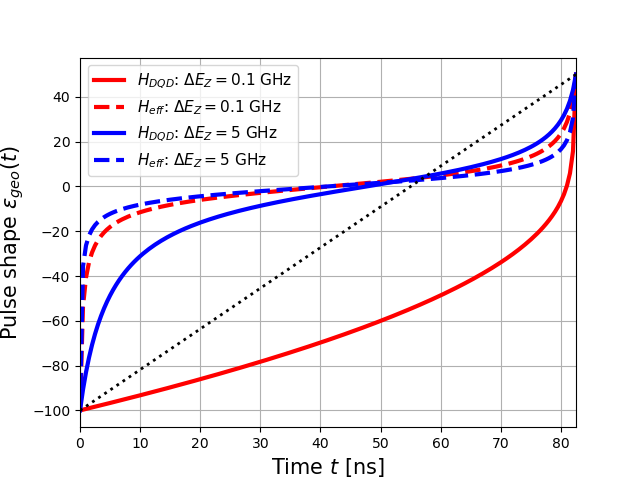}
    \caption{Pulse shaping comparison between 2d low-energy effective Hamiltonian (\ref{eqn: eff Hamiltonian SW}) and the DQD model (\ref{eqn: 3x3 Hamiltonian}). The solid lines represent the full model, whereas the dashed lines represent the effective model. The simulation was done with $\Omega=10$ GHz, and the Zeeman splitting difference was plotted as different colors. We see that the pulse shapes differ strongly in the effective regime, where $\mathcal{J}\ll 1$.}
    \label{fig: 2d vs 3d SW}
\end{figure}

As seen in the main text, when restricting to the singlet and triplet sectors for the charge configurations (2,0) and (1,1), we find the Hamiltonian (\ref{eqn: 3x3 Hamiltonian}). Given that the Zeeman splitting difference $\Delta E_Z$ is usually a much smaller energy scale than the tunnel coupling or detuning we want to find an effective 2-dimensional model. Using standard Schrieffer-Wolff transformation  we find that the effective 2d model is spanned only by the singlet sector in the two different charge states $(1,1), (2,0)$
\begin{align}
    \label{eqn: eff Hamiltonian SW}
    \hat{H}_\text{eff}(t)=\begin{pmatrix}
    -\varepsilon(t)\left(1 -\mathcal{J}^2\right)  & \Omega(1-\mathcal{J}^2/2)  \\
    \Omega(1-\mathcal{J}^2/2)  & 0
    \end{pmatrix},
\end{align}
where $\mathcal{J}^2= \Delta E_Z^2/\Omega^2$ is the expansion parameter. For this $2\times 2$ Hamiltonian we can find the fast-QUAD equation explicitly in the regime of $\mathcal{J}\ll 1$ up to second order in the Zeeman splitting difference
\begin{align}
    \frac{(1+3\mathcal{J}^2)\,\Omega^2+\varepsilon(t)^2}{(\Omega^2+\varepsilon(t)^2)^{5/2}}\left(\dv{\varepsilon}{t}\right)=\delta/\Omega.
\end{align}
In the above equation, we rescaled $\Omega \to \Omega/2$ for readability. For $\mathcal{J}=0$ we recover the fast-QUAD equation for the Landau-Zener model as seen in \cite{xuImprovingCoherentPopulation2019,fehseGeneralizedFastQuasiadiabatic2023, chenSpeedingQuantumAdiabatic2022}. The difference of pulse shapes of the effective 2d and the full 3d Hamiltonian is plotted in Figure \ref{fig: 2d vs 3d SW}. We observe that the derived pulse shapes deviate significantly between the effective and the full model. 

\section{Numerical methods}

Here we outline the numerical methods used in the main text, including the generation of the pulse shapes, the Hamiltonian, and Lindblad master equation solvers.

\subsection{Pulse shape interpolation}

First, we compute the quantum metric tensor using Eq.~\eqref{eqn: metric defintion}. Making use of the quantum adiabatic condition~\eqref{eqn: adiabatic quantum geo protocol}, we obtain a differential equation for the pulse in terms of the adiabaticity parameter $\delta$. Once, we have chosen the boundary conditions of the pulse, we can compute the adiabaticity parameter using Eq.~\eqref{eqn: adiabaticity defintion} and hence solve the differential equation for the pulse consistently between these boundary conditions. 

\subsection{Hamiltonian simulation}

For the numerical simulations of the coherent population transfer, we first generate a pulse according to the boundary conditions for initialization or readout. Next, we replace the numerically solved pulse $\varepsilon_\text{num}(t)$ in the Hamiltonian operator and then solve the Schrödinger equation (in units of $\hbar=1$)
\begin{align}
    i\dv{}{t}\ket{\psi(t)}=\hat{H}[\varepsilon_\text{num}(t)]\ket{\psi(t)}.
\end{align}

Given the initial state $\ket{\psi_0(t=0)}$, we evolve the state and project it onto the lowest energy eigenstate at the final pulse time to see whether coherent errors occurred. For that, we compute the transfer probability $p(t_\text{f})=|\braket{\psi_0(t_\text{f})}{\psi_\text{geo}(t_\text{f})}|^2$, where
\begin{align}
    \ket{\psi_\text{geo}(t_\text{f})}=\mathcal{T}\exp(-i\int_0^{t_\text{f}}\hat{H}[\varepsilon_\text{num}(t)])\ket{\psi_0(t=0)}.
\end{align}

\subsection{Lindblad simulation}

\begin{figure}
    \centering
    \includegraphics[width=0.5\textwidth]{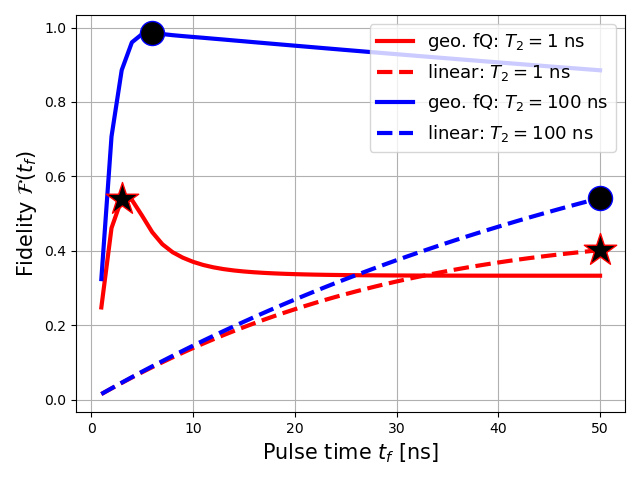}
    \caption{Example of simulated Uhlmann fidelity between the lowest energy eigenstate and the mixed state after Lindbladian evolution. The dashed lines represent the linear protocol. The stars and the dots illustrate the maximum values of the overlap for different decoherence times.}
    \label{fig: overlapExample}
\end{figure}

\begin{figure*}
    \centering
    \includegraphics[width=0.8\textwidth]{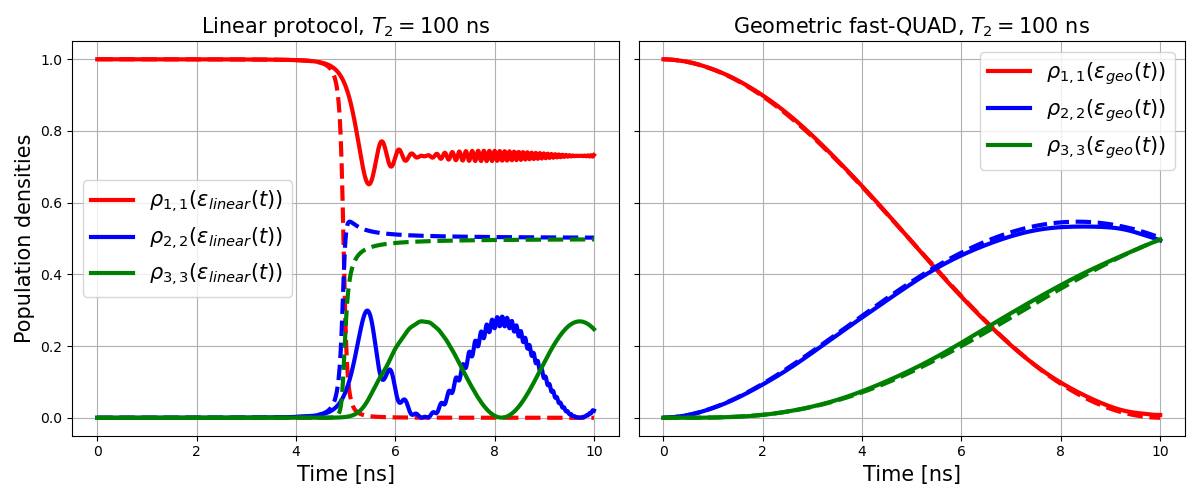}
    \includegraphics[width=0.8\textwidth]{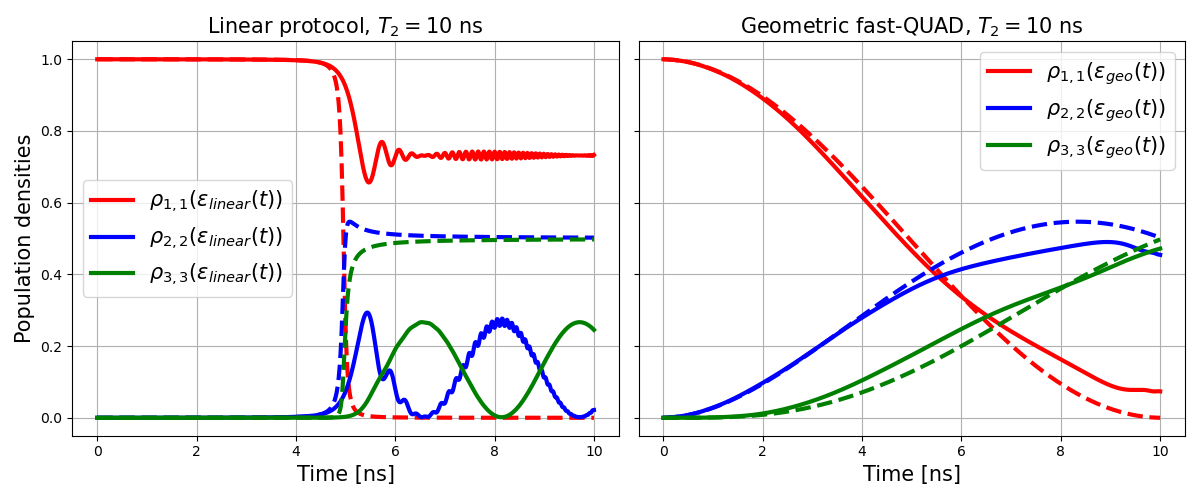}
    \includegraphics[width=0.8\textwidth]{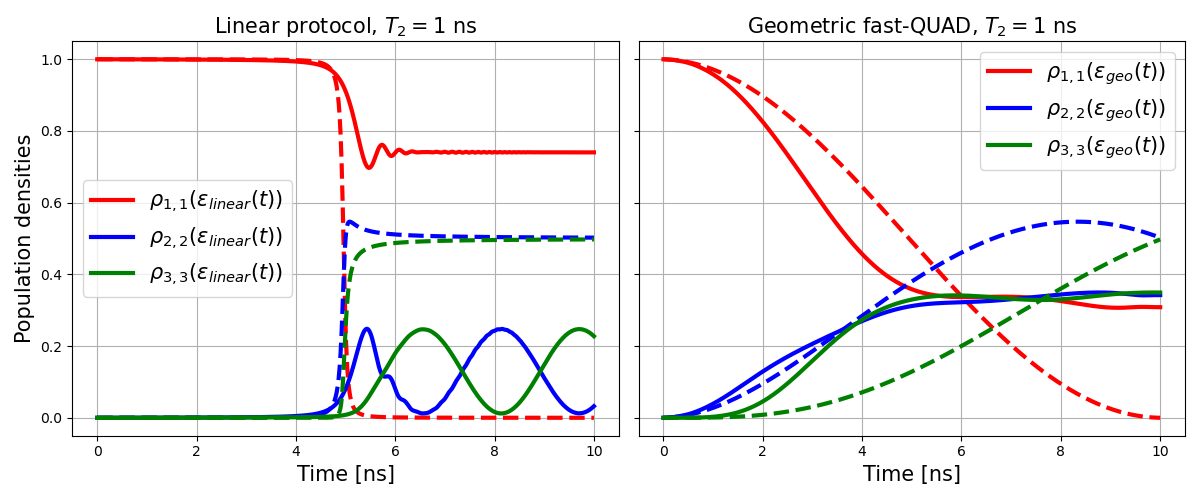}
    \caption{Comparison of the population densities during the initialization phase as a function of the linear or geometric pulse and the dephasing time $T_2$. Different colors represent different components of the population density, and dashed lines represent the noiseless evolution of the respective colored component. The parameters used are $\Tilde{U}=\unit[100]{GHz}$, $\Omega=\Delta E_Z=\unit[1]{GHz}$, and for boundary conditions $\varepsilon_0=2\Tilde{U}$ and $\varepsilon_\text{f}=0$. We note that, due to the small anti-crossing and the ultrafast pulse time ($t_\text{f}=\unit[10]{ns}$), the geometric pulse severely outperforms the linear protocol. However, only in the weak dephasing limit $T_2\gg t_\text{f}$ do we find that the geometric fast-QUAD provides a high-fidelity state transfer. The geometric protocol, due to the minimization of the Landau-Zener transitions, will remain for long times in the anti-crossing region to avoid coherent errors. Thereby, however, it will be most exposed to charge noise.}
    \label{fig: comparison population densities}
\end{figure*}

For the numerical simulations of the Lindblad master equation, we switch to the vectorized form. We choose to use vectorization by row, which means that, for instance,
\begin{align}
    \rho = \begin{pmatrix}
        a & b\\ c& d
    \end{pmatrix} \to \vket{\rho} =\text{vec}[\rho]= \begin{pmatrix}
        a \\ b\\ c\\ d
    \end{pmatrix}
\end{align}
In this notation, the Lindblad master equation can be written as a linear equation
\begin{align}
    \dv{}{t}\vket{\rho}=\hat{\mathcal{L}}\cdot \vket{\rho},
\end{align}
where the Lindbladian takes the form
\begin{multline}
    \hat{\mathcal{L}}=-i\Big(\hat{H}\otimes \hat{\mathbb{1}}-\hat{\mathbb{1}}\otimes \hat{H}^T\Big)\\
    +\sum_j\hat{L}_j\otimes \hat{L}_j^* - \frac{1}{2}\Big(\hat{L}_j^\dagger \hat{L}_j\otimes \hat{\mathbb{1}} + \hat{\mathbb{1}}\otimes [\hat{L}_j^\dagger\hat{L}_j]^T \Big).
\end{multline}
Note that the expression explicitly depends on the basis chosen, as the transpose is basis-dependent. To compute the success of the state transfer protocols we define the Uhlmann fidelity
\begin{align}
    \mathcal{F}(\rho, \sigma)=\Big( \tr \sqrt{\sqrt{\rho}\sigma \sqrt{\rho}}\Big)^2.
\end{align}
We will use this to quantify the overlap between the time-evolved pure initial state $\ket{\psi_0(0)}\approx \ket{S(2,0)}$ to the mixed state at the end of the non-unitary evolution $\rho(t_\text{f})$. In this case, the fidelity simplifies to
\begin{align}
    \mathcal{F}(\rho(t_\text{f}), \ketbra{\psi_0(t_\text{f})})=|\mel{\psi_0(t_\text{f})}{\rho(t_\text{f})}{\psi_0(t_\text{f})}|.
\end{align}
Defining $\vket{\psi_0}=\text{vec}[\ketbra{\psi_0(t_\text{f})}]$ we find that the Uhlmann fidelity reduces to
\begin{align}
    \mathcal{F}(t_\text{f})=|\vbraket{\rho(t_\text{f})}{\psi_0(t_\text{f})}|.
\end{align}
Full-time evolutions are shown in Fig.~\ref{fig: comparison population densities}. We observe that the geometric fast-QUAD provides a better overlap with the energy eigenstate for weak dephasing. Figure~\ref{fig: overlapExample} shows the procedure to obtain the optimal control simulation in Fig.~\ref{fig: infidelity nonunitary}. We start by simulating the Uhlmann fidelity for a fixed decoherence time $T_2$ for pulse times $t_\text{f}\in [0,50]\,$ns and extract the highest overlap for both the linear (dashed line) and geometric (full line) protocol. These are shown in the figure as blue circles/red stars for two exemplary decoherence times $T_2=\unit[1,100]{ns}$, respectively. Note that the geometric protocol provides a higher fidelity at shorter pulse times. From each simulation, therefore, we extract the fidelity $\mathcal{F}(t_\text{f}^*)$, the pulse time $t_\text{f}^*$ at which the highest fidelity is reached and the corresponding decoherence time $T_2$.

\clearpage

\bibliography{references.bib}

\end{document}